\documentstyle[12pt,cmb]{article}
\input epsf

\def\mathrelfun#1#2{\lower3.6pt\vbox{\baselineskip0pt\lineskip.9pt
  \ialign{$\mathsurround=0pt#1\hfil##\hfil$\crcr#2\crcr\sim\crcr}}}
\def\lesssim{\mathrel{\mathpalette\mathrelfun <}}
\def\gtrsim{\mathrel{\mathpalette\mathrelfun >}}

\def\Tell{{\ell}}

\def\cU{\Upsilon}

\def\kms{{\rm~km~s^{-1}}}

\def\mpc{{\rm~Mpc}}

\def\'{^{\prime}}
\def\avrg#1{{\langle #1 \rangle}}

\def\eps{\varepsilon}

\def\eg{{e.g., }}
\def\ie{{i.e., }}
\def\etal{{et al. }}
 
\def\etc{{etc. }}

\def\half{{\textstyle{1\over2}}}

\def\p3m{P$^3$M}

\def\spose#1{\hbox to 0pt{#1\hss}}
\def\lta{\mathrel{\spose{\lower 3pt\hbox{$\mathchar"218$}}
     \raise 2.0pt\hbox{$\mathchar"13C$}}}
\def\gta{\mathrel{\spose{\lower 3pt\hbox{$\mathchar"218$}}
     \raise 2.0pt\hbox{$\mathchar"13E$}}}
\def\ge{\mathrel{\spose{\lower 3pt\hbox{$-$}}
     \raise 2.0pt\hbox{$\mathchar"13E$}}}
\def\le{\mathrel{\spose{\lower 3pt\hbox{$-$}}
     \raise 2.0pt\hbox{$\mathchar"13C$}}}

\def\eqright{\begin{eqnarray}}
\def\endeqright{\end{eqnarray}}

\begin{document}


\heading{COSMIC PARAMETER ESTIMATION COMBINING 
SUB-DEGREE CMB EXPERIMENTS WITH COBE}

\author{J. Richard
Bond $^{1}$ \& Andrew H. Jaffe $^{1,2}$} {$^{1}$ Canadian Institute
for Theoretical Astrophysics, Toronto, Ontario, Canada.} {$^{2}$
Center for Particle Astrophysics, UC Berkeley, Berkeley CA USA.}

\begin{abstract}{\baselineskip 0.4cm 
\small
We describe the Bayesian-based signal-to-noise eigenmode method for
cosmological parameter estimation, show how it can be used to
optimally compress large CMB anisotropy data sets to manageable sizes,
and apply it to the DMR 4-year, South Pole and Saskatchewan data,
individually and in combination. A simple prior probability method is
used to include large scale structure observations. Estimates of the
Hubble parameter, vacuum energy density, baryon fraction and
primordial spectral tilt derived from the combined data are given. }
\end{abstract}

\section{Introduction} \label{sec:intro}

As CMB anisotropy experiments have gotten more ambitious, our need for
powerful statistical methods has become more urgent.  For the CMB data
sets that have been obtained up to now, including COBE \cite{dmr4} ({\it dmr4}), the
1994 South Pole data of \cite{sp94} ({\it sp94}) and the 1993-95
Saskatoon data of \cite{sk94,sk95} ({\it sk95}), which we analyze
jointly here, it is possible to do a relatively complete Bayesian
statistical analysis 
if the primary anisotropies
are assumed to be Gaussian and the non-Gaussian Galactic foregrounds are
not large. Even for these experiments, this is feasible only because of
compression, in which the data set is acted upon by linear
operators which project it onto subspaces of the full data. In the past the
linear combinations of the data were defined by what made intuitive
sense (\eg weighted sums of different frequency channels or weighted
averages of pixel separations below the beam scale), and could still
leave too many pixels to deal with in a complete analysis.
Here we use the rigorous signal-to-noise eigenmode approach to data
compression \cite{bdmr294} to reduce the sets to the manageable $\lta
1000$ important combinations. As we approach the era of megapixel data
sets promised by MAP and COBRAS/SAMBA, via the era of tens of thousands
of pixels promised by long duration balloon experiments, the question of
how to come as near to optimal compression as possible given computer
limitations becomes of paramount importance. This happy day of too many
pixels is now upon us.

The goal is to estimate the parameters $\{ y_A \}$ of a target set of
theories with angular power spectra ${\cal C}_{\Tell}$\footnote{${\cal
C}_\Tell \equiv \ell (\ell +1 ){\rm C}_\Tell/(2\pi)$, where ${\rm
C}_\Tell=\avrg{\vert a_{\ell m} \vert^2}$ is the CMB power spectrum as
usually defined and the $a_{\ell m}$ are the spherical harmonic
coefficients of the temperature fluctuations for the theoretical
signal.} by first determining the likelihood function ${\cal L}(\{ y_A
\})$ for each theory, and then comparing the likelihoods as a function
of the parameters. Using Bayes' theorem, we can write the probability
of the theoretical parameters, given the observations and the class of
theories being tested, $P(\{ y_A \} \vert {\rm OBS},\, {\rm
TH})\propto {\cal L}(\{ y_A \}) P(\{ y_A \}\vert {\rm prior})$, where
the assumed prior probability distribution $P(\{ y_A \}\vert {\rm
prior})$ can reflect {\it a priori} maximal ignorance, or take into
account constraints from other information such as large scale
structure observations. The proportionality constant is related to the
probability that the class of theories is correct given the
observations.  To give preferred values and errors for a specific
cosmological parameter of interest such as the Hubble parameter, one
often integrates (marginalizes) over the other parameters, such as the
density fluctuation power amplitude on cluster scales, $\sigma_8$, and
the primordial density fluctuation spectral index $n_s$.

In this paper, we assume the inflationary model for structure
formation with Gaussian adiabatic (scalar) density perturbations and
possibly gravitational wave (tensor) perturbations. We explore
constraints in the parameter space $\{t_0, {\rm
h},\Omega_{tot},\Omega_B{\rm h}^2 , \Omega_{vac}, \Omega_{hdm},
\Omega_{cdm}, \nu_s, \nu_t, \sigma_8 \}$. We assume reheating occurs
sufficiently late to have a negligible effect on ${\cal C}_\ell$. The
total density parameter, $\Omega_{tot}=\Omega_B+\Omega_{cdm}+
\Omega_{hdm}+\Omega_{vac}$, is expressed in terms of the densities in
baryonic cold, hot and vacuum matter, of which
$\Omega_{nr}=\Omega_B+\Omega_{cdm} + \Omega_{hdm}$ can cluster. The
age of the Universe is $t_0$ and ${\rm h}$ is the Hubble parameter in
units of $100 \kms \mpc^{-1}$; ${\rm h}t_0$ is a function of $\{
\Omega_{vac}, \Omega_{tot}\}$, so one parameter is redundant. The
scalar and tensor tilts are $\nu_s = n_s -1$ and $\nu_t$.\footnote{The
ratio of gravitational wave power to scalar adiabatic power is ${\cal
P}_{GW}/{\cal P}_{ad}\approx (-4\nu_t)/(1-\nu_t/2)$, apart from small
corrections. This determines the level of tensor anisotropies compared
with scalar. The tensor tilt is related to the deceleration parameter
$q$ of the Universe during inflation by $\nu_t/2 \approx 1+q^{-1}$
plus small corrections. Here, we take one of two cases. (no-GW case):
$\nu_t =0$, thus no gravity wave contribution (for nearly critical
acceleration, $q \approx -1$, as in natural inflation).  (GW case):
$\nu_t = \nu_s$ if the scalar tilt is negative (subcritical but nearly
uniform acceleration) and $\nu_t =0$ if the scalar tilt is positive.}
With current errors on the data, this space is too large for effective
parameter estimation. Instead we restrict our attention to various
subregions, such as $\{ \sigma_8 , n_s, {\rm h}\, | \, {\rm fixed} \,
t_0, \Omega_B {\rm h}^2 \}$, where either $\Omega_{tot}=1$ and
$\Omega_{vac}$ is a function of ${\rm h}t_0$ or $\Omega_{vac}=0$ and
$\Omega_{tot}$ is a function of ${\rm h}t_0$. The $t_0$'s we choose
are (11, 13, 15~Gyrs) with $\Omega_{vac} \ge 0$, but
$\Omega_{tot}=1$. For these cases, the ``standard'' nucleosynthesis
value $\Omega_B {\rm h}^2=0.0125$ was chosen. We constrain $0.5\le n_s
\le 1.5$, $0.5\lesssim h\le1$. For $\Omega_{tot}=1$ models, we have
roughly $\Omega_{vac} (h) \sim 0.9 [0.3({\rm h}/{\rm h}_1-1)^{0.3} +
0.7({\rm h}/{\rm h}_1-1)^{0.4}]$, where $h_1\equiv 0.5(13{\rm
Gyr}/t_0)$. A recent estimate for globular cluster ages is 
$14.6^{+1.7}_{-1.6}~{\rm Gyr}$ \cite{chaboyerGC96}. For the case
$\Omega_{vac}=0$, we have also let $\Omega_B {\rm h}^2$ vary, over the
range $0.003125 \le \Omega_B {\rm h}^2\le 0.05$, and we have also done
a limited exploration of the 13 Gyr, $\Omega_{tot}<1$, $\Omega_{vac} =
0$, space with tilt, using ${\cal C}_\ell's$ from Bond and Souradeep.
The reason for restricting the paths through parameter space is
because of the length of time required for a complete statistical
treatment of each data set per model ${\cal C}_\ell$.

\begin{figure}[htbp]
\vspace{-0.5in}
\centerline{
\epsfxsize=4.2in\epsfbox{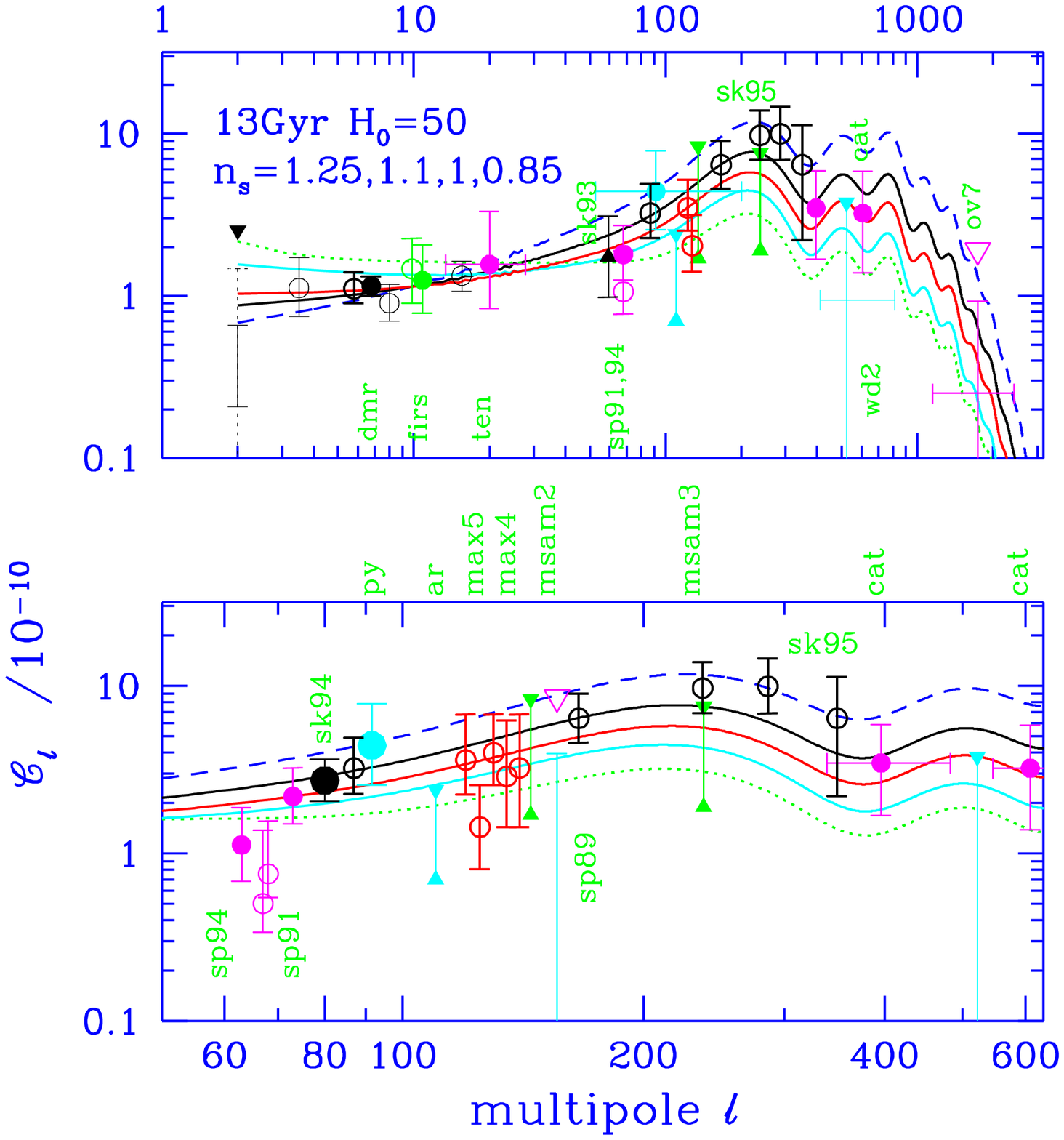}
\hspace{-0.4in}
\epsfxsize=4.2in\epsfbox{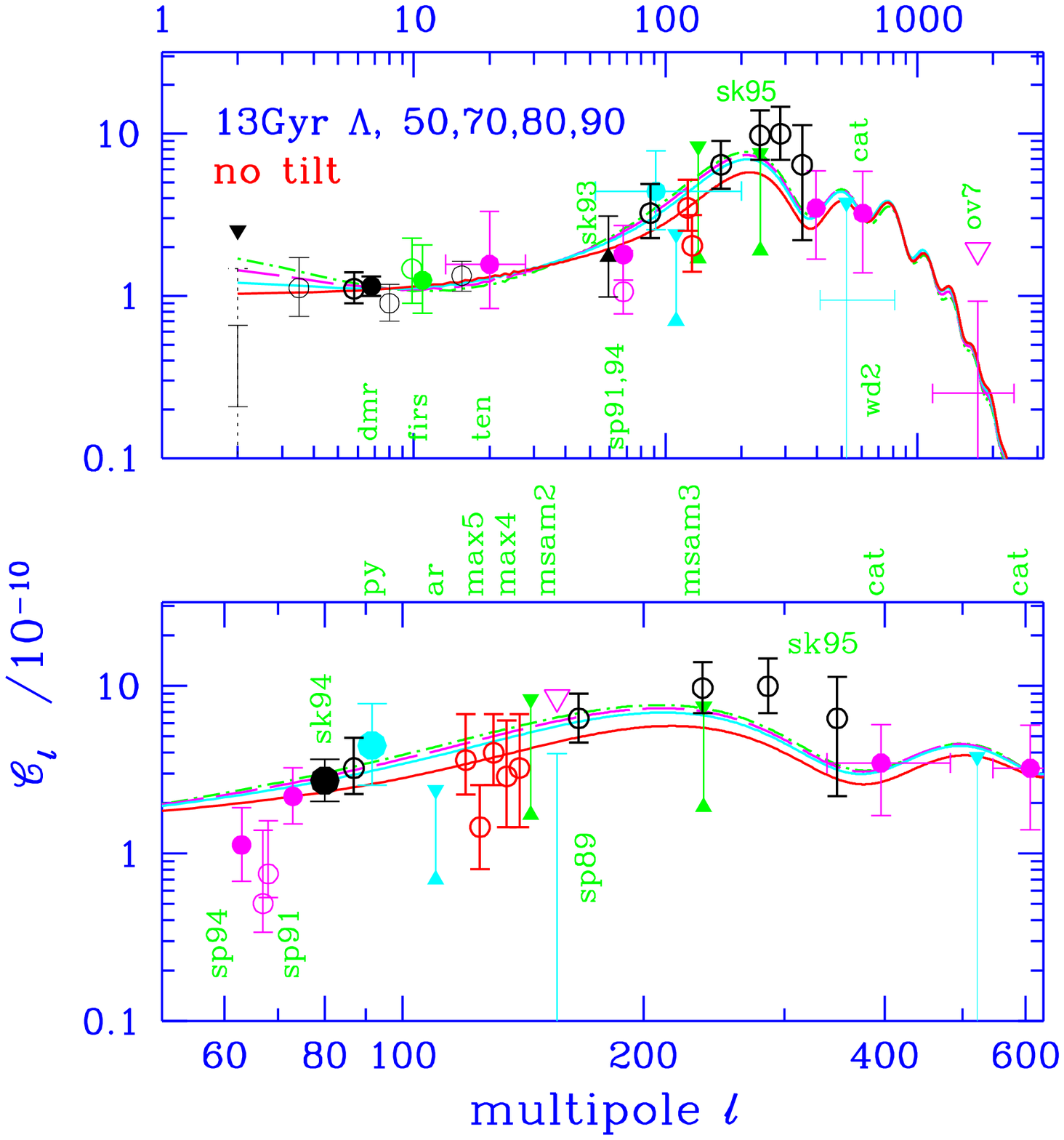}
}
\vspace{-0.2in}
    \caption{ {\small \baselineskip=0.2cm
 13 Gyr sequences,  varying $n_s$ with $\Omega_{vac} =0$, and 
varying $H_0$, hence $\Omega_{vac}$, with $n_s=1$.}}
    \label{fig:CL13Gyr}
\centerline{
\epsfxsize=4.2in\epsfbox{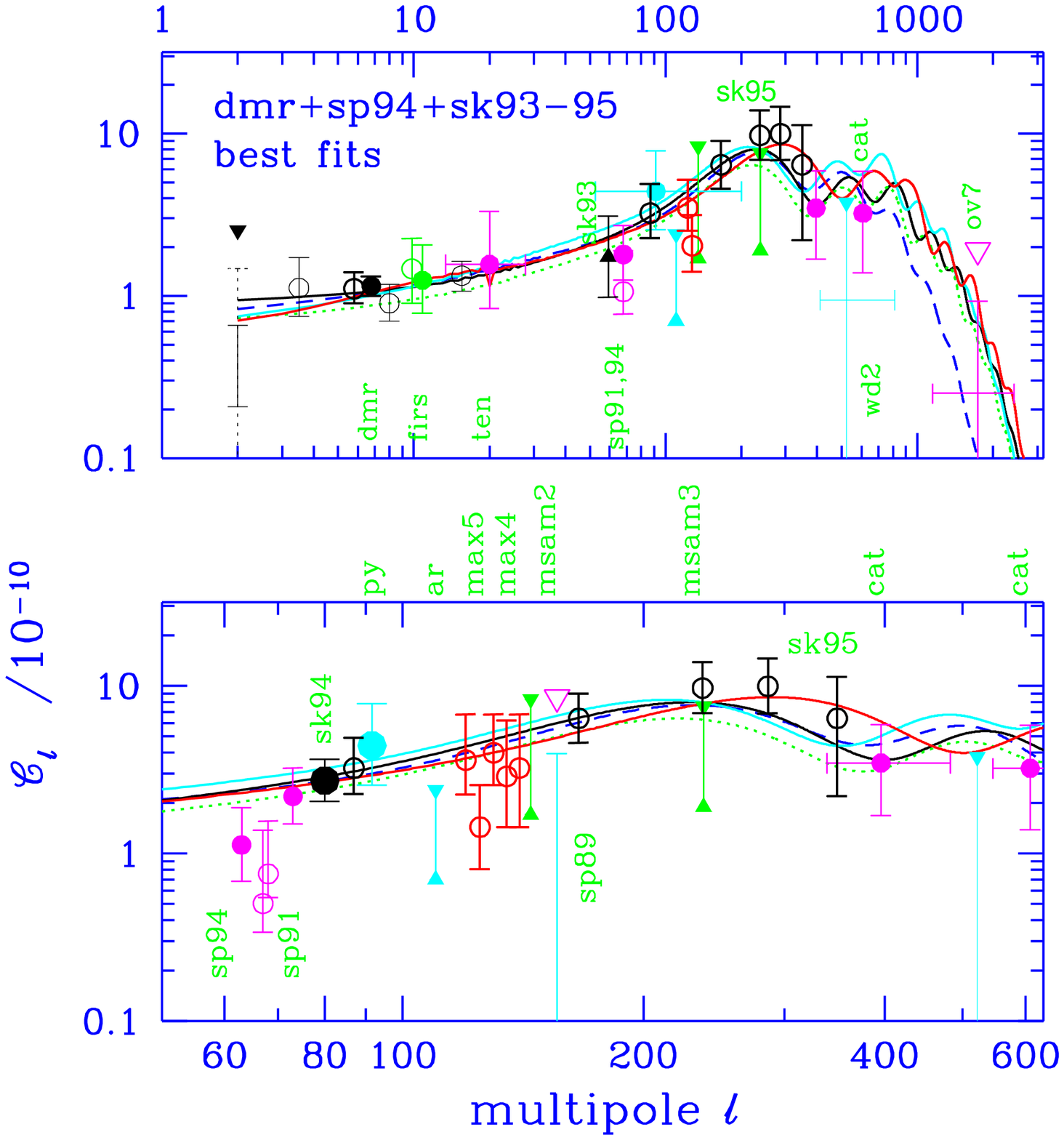}
\hspace{-0.4in}
\epsfxsize=4.2in\epsfbox{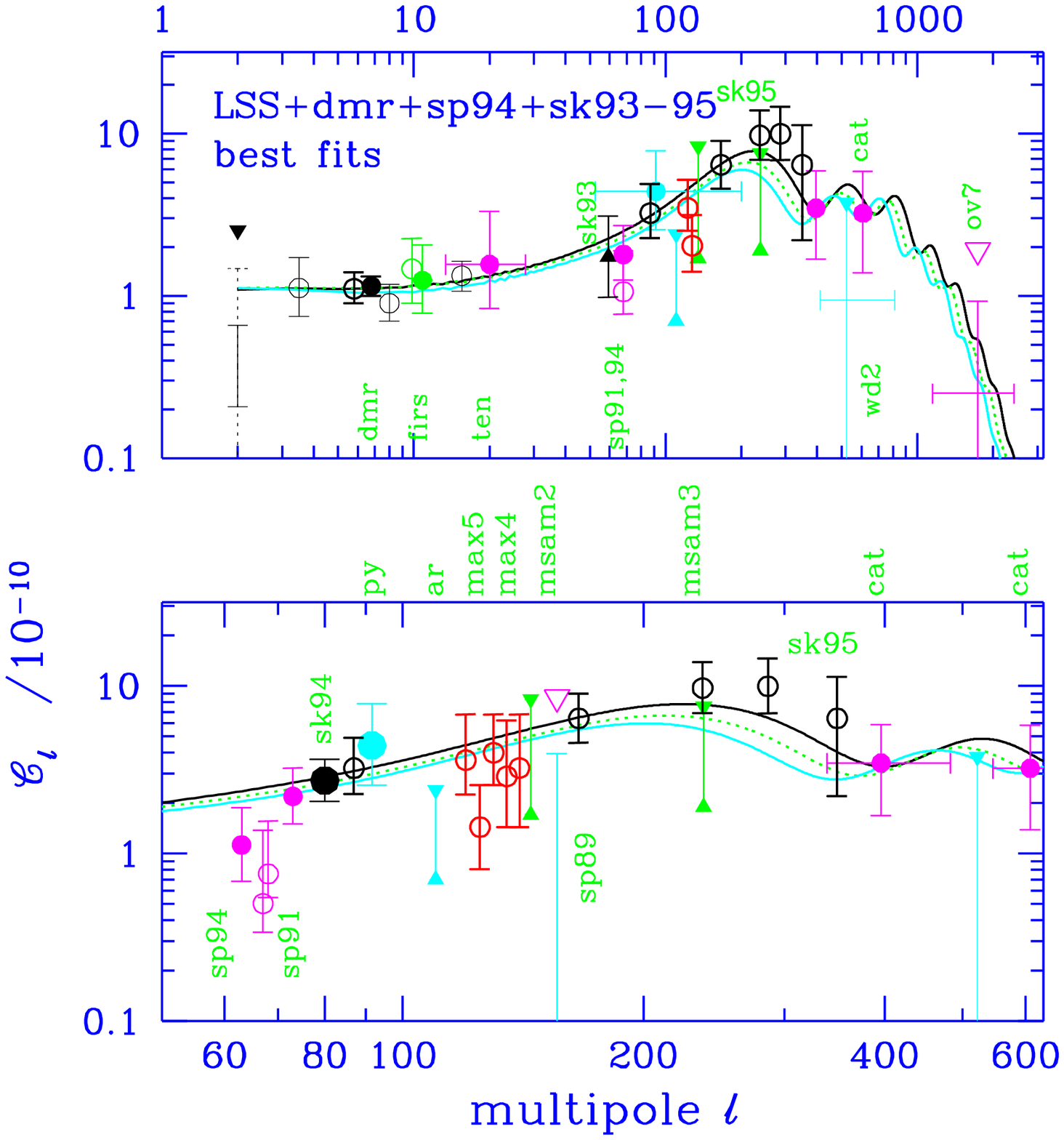}
}
\vspace{-0.25in} \caption{ {\small \baselineskip=0.0cm 
Best-fit models
using DMR+SP94+SK95 on the left, and adding LSS on the right, for 11,
13 and 15 Gyr $\Lambda$ sequences. The left also shows 13 Gyr open and
baryon sequence best fits. Parameters defining these models are given 
in Table~2. }}
\label{fig:CLbestfit}
\end{figure}

In Fig.~\ref{fig:CL13Gyr}, the bandpowers \cite{bh95} associated with
current experiments are compared with some of the ${\cal C}_\ell$'s in
the parameter space we are exploring, here the 13 Gyr, $H_0=50$, tilted
sequence with gravity waves included and the 13 Gyr $\Omega_{vac}>0$
sequence with $n_s=1$, with amplitude normalized to best-fit the
4-year DMR data. In both cases, $\Omega_B{\rm h}^2 =0.0125$. The
curves are very similar if we allow for a mix of hot and cold dark
matter with the same $\Omega_{nr}$ as these CDM models, and the other
parameters fixed. The solid dark curve is the ``standard'' untilted
CDM model. The bandpowers for the three experiments analyzed here,
{\it dmr4}, {\it sp94}, {\it sk95}, are the darker heavier data
points. Because of the differing angular scales involved we gain a
long lever arm with which we can constrain cosmological parameters
more strongly than with any individual experiment. The lower panels in
the figures are closeups of the first and second ``Doppler peak''
regions. Fig.~\ref{fig:CLbestfit} gives the best fit models, described
in \S~\ref{sec:combine}.

\clearpage

\section{Signal-to-Noise Eigenmode Method}

We are given the data in the form of a measured mean
$\overline{\Delta}_p$ of the anisotropy in the $p$th pixel, along with
the variance about the mean $\sigma_{Dp}^2$ for the measurements. In
general, there are pixel--pixel correlations in the noise, defining a
correlation matrix $C_{Dpp^\prime}$ with off-diagonal components as
well as the diagonal $\sigma_{Dp}^2=C_{Dpp}$. Also there is usually
more than one frequency channel, with the generalized pixels having
frequency as well as spatial designations. The theoretical signal also
has a correlation matrix, $C_{Tpp^\prime}$, which is a linear
combination of a product of the ${\cal C}_{\Tell}$ times a ``window
function matrix'' $W_{pp'}(\ell)$ encoding the possibly
frequency-dependent beam, the chopping strategy, sky coverage, \etc
for the experiment: $C_{Tpp^\prime} =\sum_\ell {\cal C}_\Tell
W_{pp'}(\ell)(\ell+\half )/[\ell(\ell+1)]$ (see \eg \cite{bh95}).  The
``window function'' usually reported for an experiment is
$\overline{W}_\ell=(1/N_{pix})\sum_{p=1}^{N_{pix}} W_{pp}(\ell)$.

The likelihood function is 
\begin{equation}
\ln {\cal L}(\{ y_A \} ) =
-\textstyle{1\over2}\overline{\Delta}^{\dagger}(C_n+C_T)^{-1}\overline{\Delta}
-\textstyle{1\over2} \ln {\rm det} (C_n+C_T) -N_{pix}\ln
\sqrt{2\pi} \, .   \label{eq:likeNC}
\end{equation}
Here $\dagger$ denotes transpose. The noise correlation matrix $C_n =
C_D+C_{res}$ consists of the pixel errors $C_D$ and the correlation of
any unwanted residuals $C_{res}$, such as Galactic or extragalactic
foregrounds. One can think of $C_{res}$ as increasing the noise for
selected correlation patterns in the medium. With a large enough noise
in these patterns, they are effectively projected out from the
data. 


Constraints such as averages, gradients (dipoles, quadrupoles) and known
spatial templates, which may be frequency dependent (\eg IRAS or DIRBE
combined with appropriate extrapolations) can also be modelled in the
total $\Delta_p$, as ``nuisance variables'' to be integrated
(marginalized) over. Denoting each constraint $c$ on pixel $p$ by
$\kappa_c \cU_{pc}$, where the template for constraint $c$ is $\cU_{pc}$
and the amplitude is $\kappa_c$, we need only replace
$\overline{\Delta}_p$ in eq.~(\ref{eq:likeNC}) by $\overline{\Delta}_p -
\sum_c \cU_{pc}\kappa_c $, then integrate over the amplitudes
$\kappa_c$, assuming some prior probability distribution.  This is most
easily done if we assume the $\kappa_c$ are distributed as very broad
Gaussians,  reflecting our ignorance of their values (or, if we know their
likely range, incorporating that as prior information in the Gaussian
spreads).  The integration over $\kappa_c$ then yields
eq.(\ref{eq:likeNC}) with the residual noise matrix given by
$C_{res}=\cU K \cU^{\dagger}$, where $K_{cc^\prime}=\avrg{\kappa_c
  \kappa_{c^\prime}}$ is the assumed prior variance for the constraint
amplitudes.
As the eigenvalues of $K$ become very large, the effect of the
constraint matrix is to project onto the data subspace orthogonal to
that spanned by $\cU_{pc}$. Although one can directly use the likelihood
equation in this projection limit (using $\delta^{(N_c)}(\kappa )$ for
the constraint prior), it is computationally simpler to use the Gaussian
prior.  (Taking into account constraints with amplitudes that are not
linear multipliers times the template is much more complex.)  A suitable
$C_{res}$ can also allow us to focus attention only on a specified band
in $\ell$-space for power spectrum estimation.

In practice, we do not compute the quantities
$\overline{\Delta}^\dagger\left(C_T+C_n\right)^{-1}\overline{\Delta}$
and ${\rm det}[C_T+C_n]$ directly; instead we go to a basis (\ie
linear combination of the data) in which $C_T$ and $C_n$ are diagonal.
First, we whiten the noise matrix using the nonorthogonal
transformation provided by its ``Hermitian square root,'' $C_n\to
C_n^{-1/2} C_n C_n^{-1/2}=I$; we apply the same transformation to
$C_T$ and diagonalize this in turn with the appropriate matrix of
eigenvectors, $R$: $C_T\to RC_n^{-1/2} C_T C_n^{-1/2}R^{\dagger}={\rm
diag}({\cal E}_k)$, which has units of $(S/N)^2$. We then transform
the data into the same basis, $\overline{\Delta}\to
RC_D^{-1/2}\overline{\Delta}=\overline{\xi}$, now in units of
$(S/N)$. The transformed theory matrix ${\cal E}$ still depends on the
theoretical amplitude ($\sigma_8$, etc.) as a simple multiplier, which
enables the likelihood to be easily calculated as a function of this
parameter. In the new basis, the noise and signal have diagonal
correlations and $\langle\xi_k^2\rangle=1+{\cal E}_k$, so
$\overline{\xi}_k^2$ is useful as a theory-dependent $S/N$ power
spectrum which gives a valuable picture of the data and shows how well
the target theory fares
(Fig.~\ref{fig:snmode})\cite{capripow,bdmr294,bj96}.

The modes are sorted in order of decreasing $S/N$-eigenvalues, ${\cal
  E}_{k}$, so low $k$-modes probe the theory in question best. This
expansion is a complete (unfiltered) representation of the map.  The
optimal method for data compression is to use sharp signal-to-noise
filtering, keeping only those high $S/N$ modes with $k<k_{cut}$ and
deleting low $S/N$ ones.  We also find it extremely useful to look
carefully at the power in the low $S/N$ modes to determine whether
further residuals need to be added to the generalized noise: a poor
model for the noise can give false indications of what the data is
saying and misrepresent the signal. Filtering using $S/N$-modes has a
long history in signal processing where it is called the Karhunen-Loeve
method \cite{therrien}, and it is now being widely adopted for analysis
of astronomical databases. 

For an all-sky experiment with uniform, uncorrelated
pixel variances, the eigenmodes are the spherical harmonics, and the
eigenvalues the expected coefficients $a_{\ell m}$. For a more
complicated experiment, the high-$S/N$ modes probe the peak of the
experiment's window function in $\ell$-space. Low-$S/N$ modes are more
complex. For experiments with more than one frequency channel,
differences between channels should show no CMB signal, and so the
eigenvalue should be $0$. Nearby pixels, oversampling the beam, should
also show very little signal---the smooth fall from high $S/N$ to low
traces the beam in much the same way that the window function falls as a
Gaussian $\propto\exp[-\ell^2/\ell_s^2]$ at
high $\ell$. We expect these low $S/N$ modes to be largely independent
of the theory used to calculate the appropriate $C_T$, which enables
these modes to be used as a diagnostic of both the analysis procedure
and the experiments themselves.

\section{The Experiments Analyzed}

We now discuss the anisotropy experiments we use. The six COBE/DMR
four-year maps \cite{dmr4} are first compressed into a (A+B)(31+53+90
GHz) weighted-sum map, with the customized Galactic cut advocated by
the DMR team, basically at $\pm 20^\circ$ but with extra pixels
removed in which contaminating Galactic emission is known to be high,
and with the dipole and monopole removed.  Galactic coordinate pixels
are used; slight differences arise with ecliptic coordinate
pixels. Although one can do full Bayesian analysis with the map's
$(2.6^\circ)^2$ pixels, this ``resolution 6'' pixelization of the
quadrilateralized sphere is oversampled relative to the COBE beam
size, and there is no effective loss of information if we do further
data compression by using ``resolution 5'' pixels, $(5.2^\circ)^2$
\cite{bdmr294,dmr4}. The weighted sum of channels is an exact use of
our optimal signal-to-noise compression. The resolution degradation is
not optimal but is nearly so ($C_T(\theta )$ is nearly constant for
separations $\theta$ below the scale of the beam, so adjacent pixel
differences have tiny signal but the usual data-noise). The Galactic
cut is also not optimal, but could be made so by using explicit
templates for Galactic foregrounds to include in $C_n$, as described
above. The combined effect reduces the pixel number from $6\times
6144$ to $999$. Further compression by a factor of two or so is
possible without much information loss \cite{bdmr294}. A strong
indication of the robustness of the {\it dmr} data set is the
insensitivity of the band-powers to the degree of signal-to-noise
filtering and to which frequencies are probed. For $C_{res}$, we
include templates for the monopole, dipole and quadrupole, the latter
allowing for a Galactic foreground contaminant, which we know is there
at low $\ell$ in the 31 GHz channel. The DMR data probes $\ell \sim 2
- 15$ well, with useful information out to $\sim 30$.

The {\it sp94} experiment \cite{sp94} is similar to a classic
single-differencing chopping experiment, except that differencing is
associated with the oscillation of the beam about the pixel
position. It probes $\ell\sim 30-120$.  The number of frequency
channels and spatial pixels is sufficiently small (301) that no
compression is needed: all 7 frequencies in the Ka and Q bands at
$\sim 30 $ and $\sim 40$ GHz are simultaneously analyzed. There are 14
constraints, average and gradient removals for each frequency. Taking
differences in $\Delta T/T$ in frequency at the same spatial position
is insensitive to the primary signal but has the usual pixel noise for
each channel, so $S/N$ filtering would tend to remove those modes and
strong compression would result. Because the beams do vary somewhat
with frequency, however, the compression would remove some
information, unlike for COBE.

 The {\it sk95} experiment \cite{sk95} probes a much larger band in
$\ell$-space, from $\sim 50$ to $\sim 400$. Even before the data was
delivered to us a significant amount of frequency and spatial
compression already took place. In this paper, for parameter
estimation, we use the ``CAP'' data (2016 pixels, including rebinned 
data from {\it sk94}, with 48 constraints associated with average
removals).  The SK experiment measured the temperature directly by
making slightly-curved radial scans from the North Celestial Pole
about 8 degrees in length, which covered the CAP as the earth
rotated. The data was binned in RA, but, instead of binning in
declination, it was projected in software onto what are in effect
3 to 19 beam ``chopping'' configurations.  Adding the RING data
to the CAP, involving sweeps in a ring around the NCP at $\sim
8^\circ$ brings the total to 2400 pixels, with 52 constraints.  In
\cite{bj96}, we show that the ``CAP'' and ``CAP+RING'' parameter
estimates agree to much better than ``one sigma'' even though the RING
adds substantially more data. One potential concern is that only one
HEMT band is represented in the data. We have also extensively
analyzed the SK94 data set on its own \cite{sk94}, with 1344 pixels and 28
constraints, which has only 3 to 9 beam template projections and
substantially fewer hours of integration than the 94+95 data, but the
advantage of having both Ka and Q band information so the frequency
spectrum can be checked. We agree with \cite{sk94} that the spectrum
of the SK94 3 to 9 templates is consistent with a CMB origin, and
inconsistent with likely Galactic foregrounds. (We come to the same
conclusion for the {\it sp94} data, in agreement with \cite{sp94}.)
SK95 had data only from the Q-band.

In analyzing SP94 and SK94-95, it is essential to include errors in
the overall calibration of $\Delta T$. For SK94-95 it is estimated to
be a Gaussian with standard deviation $\eps=0.14$; for SP94,
$\eps=0.10$. Let ${\cal L}_0(\sigma_8)$ denote the calculated
likelihood assuming no such errors; then the likelihood with the
uncertainty included is ${\cal L}(\sigma_8) = \int d\sigma_8^\prime
\exp[-(\sigma_8- \sigma_8^\prime )^2/(2\eps^2 \sigma_8^2)]{\cal
L}_0(\sigma_8^\prime ) $. It is unfortunate that after all of the
effort that has gone into these superb experiments, an astronomical
issue like the brightness of Cas A (for SK95) results in a
substantially poorer constraint on $\sigma_8$ than one obtains
assuming no such calibration uncertainty.

\section{Phenomenology, $S/N$ Power Spectra and Data Compression}

As we mentioned in \S~\ref{sec:intro}, we have chosen to order our
path through parameter-space using the cosmological age of the models,
$t_0$. To examine the phenomenology of the experiments we shall use a
one-parameter sequence of ${\cal C}_\ell$ shapes, with the overall
bandpower of the experiment (or the related $\sigma_8$) as another
parameter. While it was usual in the past to use a power law in ${\cal
C}_\ell$, $\sim (\ell +\half )^{\nu_{\Delta T}}$ \cite{capripow}, it is
evident from Fig.~\ref{fig:CL13Gyr} that this would be a very bad fit to the
{\it sk95} data, although it is a reasonable representation over the
limited $\ell$ range for both the {\it dmr4} and {\it sp94} data. The 
sequence we use is the first panel in
Fig.~\ref{fig:CL13Gyr}, the tilted CDM sequence for the standard CDM
model, \ie $H_0=50, \Omega=1, \Omega_B=0.05$, with $n_s$ variable.  We
use the GW case, \ie $\nu_t=n_s-1$ if $n_s <1$, $\nu_t=0$
otherwise. These models have an age of $t_0 = 13 {\rm Gyr}$. 

\begin{figure}
\vspace{-0.4in}
\centerline{
\epsfxsize=4.5in\epsfbox{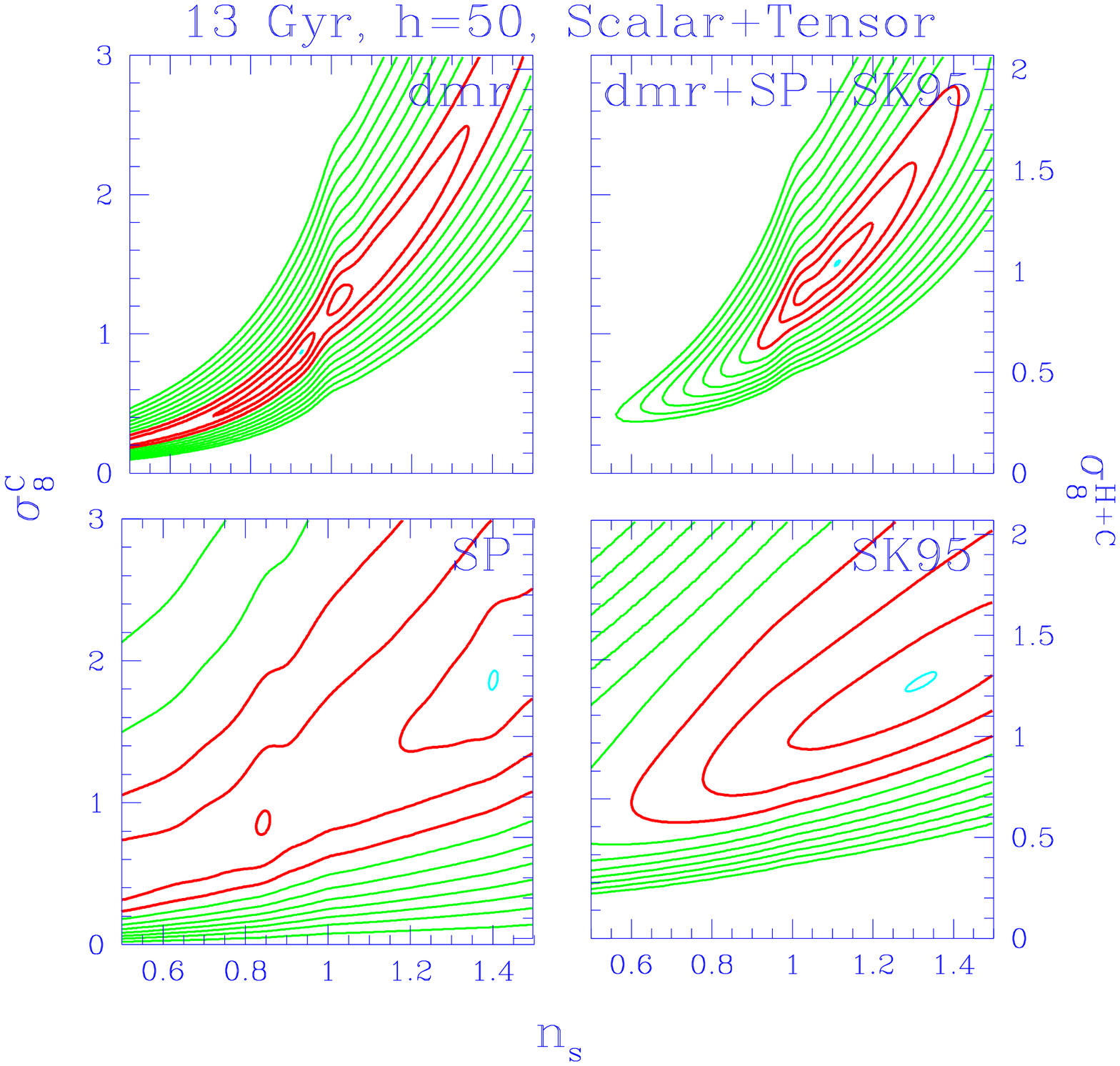}
}
\vspace{-0.2in} 
\caption{ {\small\baselineskip=0.0cm Contour plots of the likelihood
of $\sigma_8$ and $n_s$ for fixed $h=0.5$.  The contours are at ${\cal
L}$ = $\exp[-\half\{ 1,4,9,\ldots\} ]{\cal L}_{max}$ (with an extra
contour around ${\cal L}_{max}$ to show where it is). In every panel,
the lefthand $\sigma_8$ axis is for the CDM sequence, while the
righthand $\sigma_8$ axis is for a $\Omega_{hdm}=0.2\Omega_{nr}$ mixed
dark matter model. $\sigma_8$ inferred from cluster abundances is
$\sim 0.6$.}}
\label{fig:contour}
\vspace{-0.25in}
\centerline{
 \epsfxsize=4.5in\epsfbox{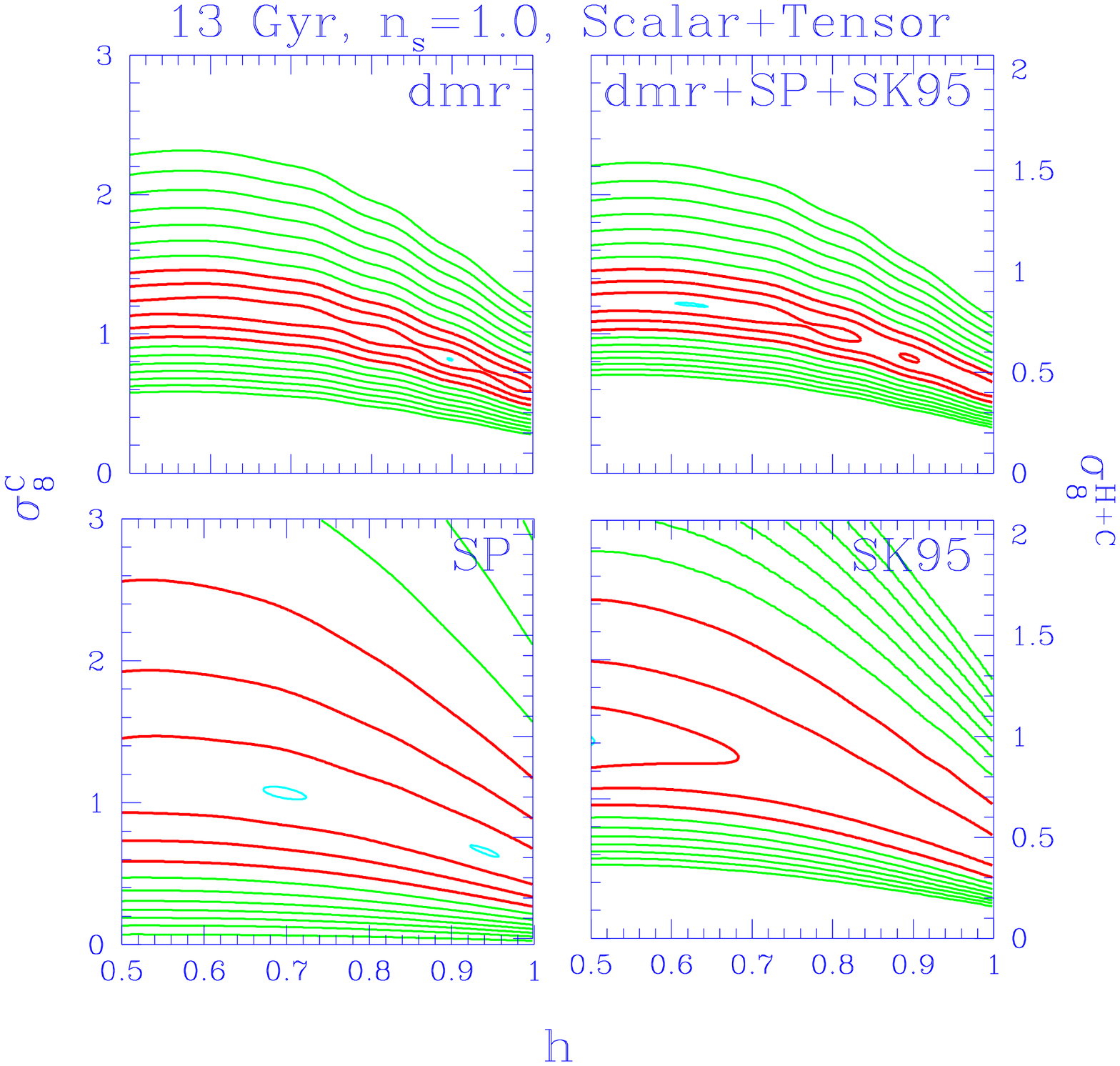}
}
\vspace{-0.2in}
    \caption{\small
Likelihood of $\sigma_8$ and ${\rm h}$ for fixed $n_s=1$. In this
case, the righthand $\sigma_8$ axis strictly applies only for the h=0.5 line.}
    \label{fig:contourh}
\end{figure}

Fig.~\ref{fig:contour} shows $1,2,3,\ldots$ sigma contours of ${\cal
L}(\sigma_8 , n_s )$, with $\nu$-sigma defined by ${\cal L}/{\cal
L}_{max} = \exp[-\nu^2/2]$. It is clear from the right hand panel that
fixing $n_s=1$ and $t_0=13$, but varying ${\rm h}$, and therefore
$\Omega_{vac}$, is not a good sequence to use for phenomenology since
there is very little difference in the ${\cal C}_\ell$'s as ${\rm h}$
varies. The Fig.~\ref{fig:contourh} ${\cal L}(\sigma_8,{\rm h})$ contour map 
shows that indeed the data does not determine the Hubble parameter
very well.

With the most recent experiments ($N_{pix} \gtrsim2000$), the computer power
required to calculate the likelihood over a sufficiently wide model
space is becoming prohibitive.  The $S/N$-eigenmodes also provide a form
of data compression which can drastically reduce the required analysis
time. By rotating to a basis in which some ``canonical theory'' with
correlation matrix $C_{T*}$ is diagonalized by the matrix $R_*$, but only
retaining some fraction of the modes, we efficiently remove parts of the
data dominated by noise (\ie modes with very low $S/N\ll1$), but
retain the Gaussian character of the likelihood for the remaining modes.
For other theories, the transformed theory matrix $(R_*C_n^{-1/2} C_T
C_n^{-1/2}R_*^{\dagger})$ will no longer be diagonal, so the full matrix
calculation must still be performed, but now on the smaller space of
observations restricted to the modes with the highest ``canonical''
$S/N$.  Moreover, for these theories, the $S/N$-modes will be somewhat
different, so the compression will not be as efficient (\ie we will
have thrown out a bit ``more signal and less noise'').  Still, we have
achieved compression as good as 90\% for experiments (like SK94) with
two channels and 65\% for the full SK94-95 dataset, which has already
been re-binned to remove some of the redundancy in beam oversampling and
channel-to-channel differences.  Because the cost of the matrix
calculations involved scales as $N^3$, these result in significant
speedups: from $\sim 1$~hour to $\sim 10$~minutes per point in parameter
space for {\it sk95} (which is actually significantly worse than the expected
$(1-0.65)^3\simeq1/25$ speedup due to overhead). 

\begin{figure}
\vspace{-0.4in}
\centerline{
\epsfxsize=4.5in
\epsfbox{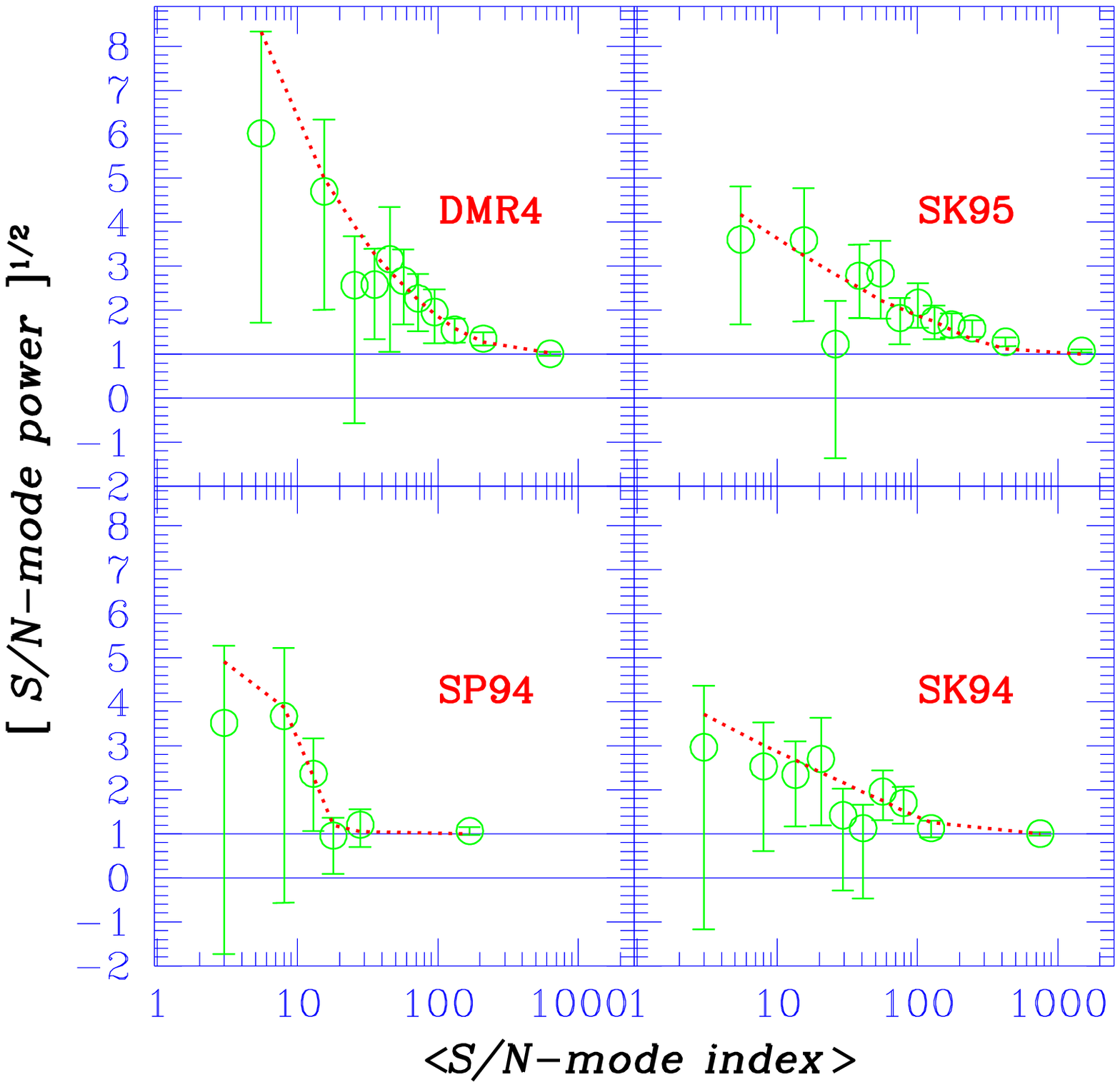}
}
\vsize=-0.4in
    \caption{\small
Observed and theoretical $S/N$-spectra with $1\sigma$
errors (including pixel noise and cosmic variance), using ``standard''
CDM, with $\sigma_8=1.2$, the DMR4 value, for the theory.}
    \label{fig:snmode}
\centerline{
\epsfxsize=4.5in
\epsfbox{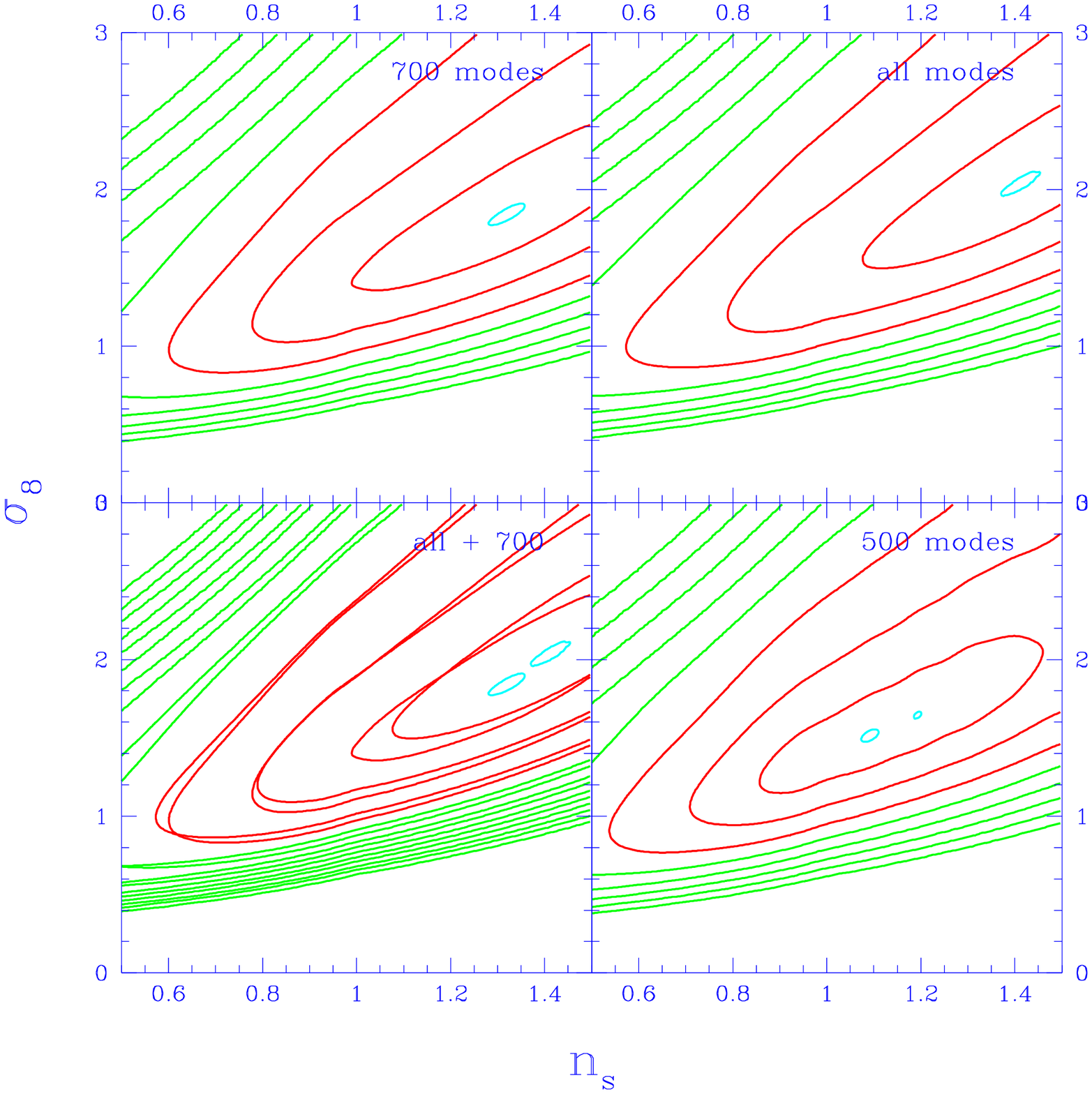}
}
\vsize=-0.4in
   \caption{{\small\baselineskip=0.2cm 
How $\sigma_8$-$n_s$ likelihood
       contours  change with
     the SK95 $S/N$ cut:  for 500, 700 and all ($\sim 2000$) modes.}}
    \label{fig:compress}
\end{figure}

In Fig.~\ref{fig:compress}, we show $1,2,3,\ldots$ sigma likelihood
contours for the SK94-95 CAP dataset, as in Fig.~\ref{fig:contour}.
The lower left panel superposes the contours of the $k_{cut}=700$ case
upon those with all 2016 modes included. The similarity of the
contours shows that both the amplitude and the index determinations
are not compromised by $S/N$ cuts. For the $k_{cut}=500$ case,
contours $\ge 2$ are very similar as well. Thus we can achieve
significant degrees of compression without loss of information. In the
following, we apply no data compression to the analysis of the DMR and
SP data, but for SK95 we present results using the top 700 modes from
the canonical standard CDM theory, the $n_s=1$ model in the
$t_0=13$~Gyr sequence.

The reason the compression works can be understood by examining the
$S/N$ power spectra, shown in Fig.~\ref{fig:snmode}.  The curve is the
theoretical spectrum $1+{\cal E}_k$ given by the eigenmodes, for a
``standard CDM'' model with amplitude $\sigma_8=1.2$, the value
suggested by COBE. The points are the observations $\overline{\xi}_k^2$, with the
same binning as the theory curve.  (The bins require a certain
signal-to-noise when summed, but a minimum number are required to
define a bin so that the error bars are not too large.)
The error bars contain both variances associated with the pixel noise
and with the theoretical cosmic variance (noise-noise, noise-signal
and signal-signal terms). To be a good fit to the data, the error
bars should pass through the theory curve. After the top few hundred
modes, the eigenvalues have $S/N\ll1$ so we do not expect them to
contribute significantly to the likelihood. We emphasize that it is
legitimate to use any mode subset: the relative likelihoods we obtain
will tell us which theory is preferred for those modes.  It is just
that we do not want to build any prior prejudice for a theory by
compressing the data in a way which may be biased in its favour over
the other theories we are testing. Thus we choose to go far into the
$S/N$ tail, retaining 700 modes. The $S/N$ mode formalism
also can be used to design experiments to discriminate particular
theories (\eg Knox, these Proceedings).

\section{Combining Experiments and Parameter Estimation} \label{sec:combine}

Combining experiments to get a total likelihood is straightforward. If
the pixels are uncorrelated, either because they overlap little on the
sky or in $\ell$, we only need to
multiply the individual likelihoods together. This is the case for
COBE/DMR, SP94, SK94-95. If there is significant overlap, then the
experiments should ideally be combined and considered to be one larger
experiment, with $C_T$ connecting the pixels in one experiment with
the pixels in the other, although the cross-pixel $C_D$ will be zero.
We have applied this to the SK95+MSAM dataset, in joint work with
Charbonneau and Knox, but will not describe it here. 

The upper right panel of Fig.~\ref{fig:contour} shows the likelihood
for the $H_0=50$, GW, 13 Gyr, tilted sequence. Each experiment
individually constrains the amplitude, $\sigma_8$, better than the
shape, $n_s$: the window functions cover a narrow range of
$\ell$. Note that the SK experiment does better than DMR at
determining the slope.  The calibration uncertainty for SK95 is the
reason that $\sigma_8$ is not more tightly
constrained. Fig.~\ref{fig:contour} shows the advantage offered by
combining the results of different experiments: the long baseline in
$\ell$ helps considerably in localizing the $n_s$ contours. In
Fig.~\ref{fig:contourh}, we see that the DMR data does not restrict
the value of ${\rm h}$, and thus not of $\Omega_{vac}$, whereas the SK
data does, yet the combined data focusses the $\sigma_8$
determination, but not the ${\rm h}$ determination. The SK95 data
prefers more power than is predicted by the DMR data for ``standard''
models. Thus, the $n_s=1$, $H_0=50$, $\Omega_B {\rm h}^2 = 0.0125$
model has $\sigma_8 = 1.20 \pm 0.08$ for DMR, while for SK95 it is
$\sigma_8 = 1.48^{+0.26}_{-0.20}$. Increasing $\Omega_B {\rm h}^2 $ to
0.2 brings them closer into line, $\sigma_8 = 0.89 \pm 0.06$ for DMR,
while for SK95 it is $\sigma_8 = 0.82^{+0.15}_{-0.11}$; and this model
is much preferred statistically to the 0.0125 one.

We could repeat the contour maps for the 11 and 15 Gyr vacuum
sequences, for the fixed $t_0$ open models, and for the variable
$\Omega_B$ sequence, but it is more concise to quote single numbers,
our estimates of the individual cosmological parameters. To that end,
we marginalize over the other parameters in the sequence, assuming a
prior probability for the parameters.  If it is uniform we get the
results in the left columns of Table~\ref{tab:results}. The idea that
first motivated this project was that the SK94 and SP94 data looked
sufficiently robust to return to the multiresolution approach
combining experiments to get best possible constraints, \eg
\cite{bh95}, and this would significantly improve the COBE-only errors
on $n_s$.  The SK94+SP94+DMR4 column is the culmination of that
effort. However, the SK95 data took us to significantly higher $\ell$
and the promise of greater discrimination among models based on how
they rise to the Doppler peak. Notice the rather large shift in $n_s$
when we pass from the 1300 pixel 2-channel SK94 data, which had
chopping templates from 3 to 9, to the SK95 set, which had 10-19
projections as well as much more 3-9 data, but only for the Q-band. A
worry is that non-cosmic signals at high $\ell$ might be contaminating
the 10-19 template projections.

\begin{table}[htbp]
  \begin{center}
    \leavevmode
\begin{tabular}{c|c|ccc|cc}
\hline
\multicolumn{2}{c|}{\bf case} & {\bf DMR4} &  {\bf SK94+SP94}&{\bf
SK94-95}& {\bf LSS}+DMR4 & {\bf LSS}+DMR4   \\
 \multicolumn{2}{c|}{} & &  +DMR4&+SP94+DMR4 &  & +SK94-95+SP94 \\
\hline
\multicolumn{7}{l}{13 Gyr $\Omega_{vac}$ sequence, with GW} \\
\hline
$H_0$=50 & $n_s$ & $1.02^{+.23}_{-.18}$ & $0.95^{+.05}_{-.08}$ &
$1.12^{+.11}_{-.09}$ & $0.76^{+.03}_{-.03}$ &$ 0.85^{+.02}_{-.02}$ \\
$H_0$=70 & $n_s$ & $1.12^{+.26}_{-.24}$ & $0.92^{+.05}_{-.05}$ &
$1.12^{+.09}_{-.12}$ & $0.99^{+.03}_{-.02}$ & $0.99^{+.03}_{-.02}$ \\
All $H_0$ & $n_s$ &   &  & $1.11^{+.11}_{-.10}$ & $1.02^{+.40}_{-.06}$  & $1.07^{+.13}_{-.08}$ \\
$n_s$=1 & $H_0$ &$ <90$ & $<65$ & $<65$ & $68\pm 3$ & $ 67 \pm 3$ ($\Omega_{vac}\approx 0.60$)\\
All $n_s$ & $H_0$ &   &  & $<76$ & $<68$& $70 \pm 4$ ($\Omega_{vac}\approx 0.66$) \\ 
\hline
\multicolumn{7}{l}{15 Gyr $\Omega_{vac}$ sequence, with GW}\\
\hline
$H_0$=43 & $n_s$ & $1.02^{+.22}_{-.18}$ & $0.93^{+.05}_{-.06}$ &
$1.04^{+.13}_{-.06}$ & $0.84^{+.03}_{-.03}$ &$0.91^{+.03}_{-.04}$  \\
$H_0$=70 & $n_s$ & $1.28^{+.21}_{-.33}$ & $0.89^{+.08}_{-.08}$ &
$1.02^{+.16}_{-.06}$ & $1.25^{+.04}_{-.04}$ & $1.26^{+.03}_{-.04}$ \\
All $H_0$ & $n_s$ &   &  & $1.05^{+.12}_{-.08}$ & $1.17^{+.15}_{-.18}$ & $1.03^{+.14}_{-.04}$ \\
$n_s$=1 & $H_0$ &$ <76$ & $<51$ & $<61$ & $54\pm 3$ & $ 53 \pm 2$ ($\Omega_{vac}\approx 0.50$)\\
All $n_s$ & $H_0$  &   &  & $<63$ & $<57$ & $ 55 \pm 4$
($\Omega_{vac}\approx 0.52$)\\ 
\end{tabular}
\caption{\small Some results for $n_s$ and $H_0$ from our analysis for
various combinations of experiments as indicated. Upper limits and
error-bars are ``1-$\sigma$'' ones determined using likelihood ratios
${\cal L}=e^{-1/2}{\cal L}_{max}$. In rows 1, 2 and 4 for each case,
$H_0$ or $n_s$ is fixed and only $\sigma_8$ is marginalized. The other
two marginalize over $H_0$ or $n_s$ as well as $\sigma_8$. In all
cases, we have fixed $t_0=13$~Gyr, $\Omega_B{\rm h}^2=0.0125$,
$\Omega_{tot}=1$ and included tensor modes for $n_s<1$.  Note that
whether the small angle experiments are added to LSS+DMR4 or not has
little impact on the $n_s$ ($H_0$) estimation if $H_0$ ($n_s$) is fixed.}
\label{tab:results}
  \end{center}
\end{table}

\begin{table}[htbp]
  \begin{center}
    \leavevmode
\begin{tabular}{c|ccccc}
\hline
\multicolumn{6}{l}{DMR4+SP94+SK94-95 BEST FIT MODELS, with GW} \\
\hline
{\bf case}  & $H_0$  &  $n_s$ & $\Omega_{vac}$ & $\sigma_8$ & \\
\hline
13 Gyr $B$ & 50 & $1.15$ & 0 & $1.8$ (1.24) & ($\Omega_B=0.013$) \\
13 Gyr $B$ & 50 & $1.0{\bf F}$ & 0 & $0.9$ (0.62) & ($\Omega_B=0.17$) \\
15 Gyr $\Lambda$ & 43 & $1.05$ & 0 & $1.1$ (0.76) & ($\Omega_B=0.068$) \\
13 Gyr $\Lambda$ & 50 & $1.15$ & 0 & $1.3$ (0.90) & ($\Omega_B=0.05$) \\
13 Gyr $\Lambda$ & 60 & $1.0{\bf F}$ & 0.43 & $1.22$  & ($\Omega_B=0.035$) \\
11 Gyr $\Lambda$ & 59 & $1.2$ & 0 & $2.2$ (1.54)  & ($\Omega_B=0.036$) \\ 
13 Gyr OPEN & 55 & $1.05$ & $\Omega_{tot}$=0.60 & 1.0 & ($\Omega_B=0.041$)  \\
\hline
\multicolumn{6}{l}{LSS + DMR4+SP94+SK94-95 BEST FIT MODELS, with GW} \\
\hline
15 Gyr $\Lambda$ & 55 & $1.0$ & 0.52 & $0.97$  &  (best fit) \\
13 Gyr $\Lambda$ & 65 & $1.0$ & 0.56 & $1.17$  & ($1\sigma$ down)  \\
11 Gyr $\Lambda$ & 85 & $1.05$ & 0.69 & $1.49$ & ($2\sigma$ down)  \\ 
\hline
\end{tabular}
\caption{``Best fit'' models for various regions in the scanned
parameter space. The top panel considers the combined likelihoods from
the CMB experiments.  In the $\sigma_8$ column, the numbers in
parentheses are the equivalent values for a cold+hot dark matter
universe with $\Omega_{hdm}=0.2$. For the $\Lambda$ cases, $\Omega_B$
is not varied separately, but fixed at $\Omega_B{\rm h}^2=0.125$. The
bottom panel adds in the prior information from large-scale structure
and lists the best fit model for each age, as well as their relative
likelihoods. ${\bf F}$ means the parameter value was forced to be the
one indicated. }
\label{tab:bestfits}
\end{center}
\end{table}

We now wish to add some large scale structure constraints, by
constructing prior probabilities that roughly correspond to the
restrictions arising from observations of galaxy clustering and
cluster abundances.\footnote{We can also augment the prior probability
with likelihoods for CMB experiments that we have neither the data nor
the time to analyze properly, by using the quoted flat bandpower
alone, along with the appropriate window function. Many groups are now
using this technique exclusively to estimate CMB parameters; in joint
work with Knox we show how well this extreme form of compression
works.}  The linear power spectrum for density fluctuations is often
characterized by the shape parameter, $\Gamma \approx \Omega_{nr}{\rm
h} \, e^{-[\Omega_B(1+\Omega_{nr}^{-1}(2{\rm h})^{1/2}) -0.06]}$,
which is 0.48 for the standard CDM model. Here
$\Omega_{nr}=\Omega_B+\Omega_{cdm} + \Omega_{hdm}$.  Assuming a linear
bias model for how the galaxy distribution is amplified over the mass
distribution, the clustering data implies $0.15 \lta \Gamma +\nu_s/2
\lta 0.3$. The abundance as a function of X-ray temperature also
heavily constrains $\sigma_8$. Values from 0.5 up to 0.7 are obtained
for $\Omega_{nr}=1$ CDM-like models.  For $\Omega_{vac} > 0$, the
value is higher, scaling roughly as $\Omega_{nr}^{-0.56}$. There are
also many estimates of the combination $\sigma_8 \Omega_{nr}^{0.56}$
that are obtained by relating the galaxy flow field to the galaxy
density field inferred from redshift surveys, which all take the form
$[b_g\sigma_8] \, \beta_g$, where $b_g$ is the galaxy biasing factor
and $\beta_g$ is a number whose value depends upon data set and
analysis procedure: \cite{strausswillick} give $0.64 \pm 0.27$ for an
average of a number of estimates in the literature, and $ 0.55 \pm
0.10$ for a determination using a maximum likelihood technique for the
IRAS survey and the Mark III velocity field data set, while a higher
($\sim 0.7$) number is obtained using POTENT on this data set. It is
usual to take $b_g \approx \sigma_8^{-1}$ for galaxies, which gives a
$\sigma_8$ consistent with the cluster value, but $b_g$ can depend
upon the galaxy types being probed, upon scale, and could be bigger or
smaller than $\sigma_8^{-1}$.

We want to choose priors for $\sigma_8$, $\Omega_{vac}$ and $\nu_s$
that reflect these LSS ranges, but we certainly don't want to be too
miserly in our choice of allowed ranges. A straight Gaussian tends to
be overly supportive of the mean, while a tophat error has no
probability in the wings. Using priors which convolve a Gaussian with
a tophat and have different upper and lower errors give us the
flexibility we require. It is similar to specifying both a statistical
and a systematic error. For the exercise shown in the tables, we
required that $\Gamma +\nu_s /2$ be $0.22^{+0.07+0.08}_{-0.04-0.07}$
and $\sigma_8 \Omega_{nr}^{0.56} $ be
$0.65^{+0.02+0.15}_{-0.02-0.08}$. The latter has a high probability at
0.55, but little at 0.50 (although some authors actually prefer this
value). Sample LSS+CMB numbers are given in Table~\ref{tab:results}.

The tiny error bars when LSS constraints are added to the CMB data are
amusing, but are far from definitive at this stage. The reason the
errors are small is typically that the CMB data pushes for a
likelihood peaked at high $\sigma_8 \gta 1$, and this multiplies the
LSS likelihood peaking at 0.6 or so. The product of the two has a
narrow peak but also a small likelihood. This asymmetry is not as
pronounced for the hot/cold hybrid models. 

Table~\ref{tab:bestfits} gives the parameters for the best fits to the
data for the various cases. The associated ${\cal C}_\ell$'s are shown
in Fig.~\ref{fig:CLbestfit}. Note that the models which best fit the
CMB data for a given age often have positive tilts. While positive
scalar tilt is possible in inflation models, it requires special
constructs in the inflaton potential in a region corresponding to just
where we can observe it with the CMB. More likely are negative
tilts. If we restrict our attention to these (\eg second row), then
the best fit for the $\Omega_B$ sequence (13 Gyr, $H_0=50$) is $n_s
\approx 1 $ and $\Omega_B \approx 0.17$, high not low. In the lower
LSS part of Table~\ref{tab:bestfits}, the 13 Gyr best fit is one sigma down from
the 15 Gyr best fit, and the 11 Gyr is two sigma down.

The analysis shows that $n_s$ lies close to the value predicted by
inflation. The $H_0$ limits are suggestive, but better CMB data is
needed to strengthen the constraint to usable values. Of course when
the LSS data are included, $\Omega_{vac} >0$ is suggested for $n_s=1$
CDM models, though it is not needed for hot/cold hybrids with
$\Omega_{hdm}=0.2 \Omega_{nr}$. Table~\ref{tab:results} shows adding
SK95 and SP94 to LSS and DMR4 does not add much further
discrimination, but this should change dramatically in the next few
years, with the advent of long duration balloon experiments,
interferometers, MAP and COBRAS/SAMBA.

We would like to thank Barth Netterfield and Lyman Page for helping us
understand how to model their data and Lloyd Knox for useful
conversations.

\small
\def\prd{{Phys.~Rev.~D}}
\def\prl{{Phys.~Rev.~Lett.}}
\def\apj{{Ap.~J.}}
\def\apjl{{Ap.~J.~Lett.}}
\def\apjsuppl{{Ap.~J.~Supp.}}
\def\mnras{{M.N.R.A.S.}}


\begin{thebibliography}{99}{
\parskip=0.0cm 
\baselineskip 0.0cm

\bibitem{dmr4} C. Bennett \etal, 1996, \apjl {\bf 464}, L1; and
4-year DMR references therein.





\bibitem{sp94} J.O. Gundersen \etal, 1995, \apjl, {\bf 443}, L57 {\it sp94}.

\bibitem{sk94} C.B. Netterfield, N. Jarosik, L. Page \& D. Wilkinson, 1995, 
\apjl\ {\bf 455}, L69. {\it sk94}.

\bibitem{sk95} C.B. Netterfield, M.J. Devlin, N. Jarosik, L. Page \&
E.J. Wollack, 1996, \apj, submitted {\it sk95}.


\bibitem{bdmr294}
J.R. Bond, 1994, \prl\ {\bf 74}, 4369.

\bibitem{chaboyerGC96}
B. Chaboyer \etal, 1996, Science, in press.  

\bibitem{bh95} Bond, J.R. 1996, {\it Theory and Observations of the
Cosmic Background Radiation}, in ``Cosmology and Large Scale
Structure'', Les Houches Session LX, August 1993, ed. R. Schaeffer,
Elsevier Science Press. 

\bibitem{capripow} J.R. Bond, 1995, 
{Astrophys. Lett. \& Comm.}, {\bf 32}, 63.

\bibitem{bj96}
Bond, J.R. \&  Jaffe, A., 1996, CITA preprint.

\bibitem{therrien}
C.W. Therrien 1992,
{ Discrete Random Signals in Statistical Signal Processing},
ISPN0-13-852112-3 (Prentice Hall).

\bibitem{strausswillick}
M. Strauss, \& J. Willick, J.  1995, {\it Phys. Rep.} {\bf 261}, 271 





}\end{thebibliography}
\end{document}